\documentclass[a4paper,11pt]{article}
\pdfoutput=1 

\usepackage{jheppub} 

\usepackage[T1]{fontenc} 
\usepackage{slashed}

\title{\boldmath The D3-probe-D7 brane holographic fractional topological insulator}

\author[a]{Charlotte Kristjansen}
\author[b]{and Gordon W.\ Semenoff$\,$}

\affiliation[a]{Niels Bohr Institute, Copenhagen University,\\ Blegdamsvej 17, 2100 Copenhagen \O, Denmark}
\affiliation[b]{Department of Physics and Astronomy, University of British Columbia,\\ 6224 Agricultural Road, Vancouver, British Columbia, Canada V6T 1Z1}

\emailAdd{kristjan@nbi.ku.dk}
\emailAdd{gordonws@phas.ubc.ca}

\abstract{
The D3-probe-D7 brane system, oriented so as to have 2+1-dimensional Poincare symmetry, is argued
to be the holographic representation of a strongly correlated fractional topological insulator which exhibits 
a zero-field quantized Hall effect with
half-units of Hall conductivity. The phase diagram of the system with charge density and external magnetic
field is found and, as well as charge gapped quantum Hall states, it exhibits metallic and semi-metallic
phases with interesting behaviours. The   relationship of this to other models of fractional topological insulators
is discussed. }

\begin{document} 
\maketitle
\flushbottom

\section{Introduction and summary}

\subsection{Summary of results}

One of the prototypical examples of a topological insulator \cite{moore}-\cite{Ando:2013bqa} 
was suggested by Haldane \cite{Haldane:1988zza} who took
graphene, with its emergent relativistic fermions \cite{Semenoff:1984dq}, and added a parity and time-reversal violating mass term.
The result is an insulator which would exhibit a ``zero-field Hall effect'',  where it has a single unit of integer
quantum Hall conductivity
(per fermion  spin state), even in the absence of a magnetic field, and indeed, even in the absence of a charge
density.  As well as zero field Hall effect, his time reversal violating insulator has many interesting features, 
an example being that, in the appropriate geometry, 
it could furnish a unique physical example of a magnetically charged image particle \cite{Qi,Estes:2012nx}.  
 
In this paper, we will discuss the string theory dual of the highly correlated, 
strong coupling limit of such a topological insulator, the D3-D7 brane
system.  Like Haldane's model, this system has a half-unit of Hall conductivity per species of fermion.  
Haldane's model is based on 
a lattice and the Nielsen-Ninomiya theorem \cite{Nielsen:1980rz,Nielsen:1981xu}
 requires  fermion doubling, resulting in two species of fermions 
and producing a single unit of Hall conductivity per spin state.  The latter is
 in agreement with the topological TKNN integer \cite{Thouless:1982zz} which computes
 the Hall conductivity of a lattice fermion. 
In the case
of the D7 brane, there is nothing apparently wrong with a system with a single brane and a single species of fermions.
As a result, the Hall conductivity can come in half-integer units, and we therefore refer to it as a fractional topological insulator.

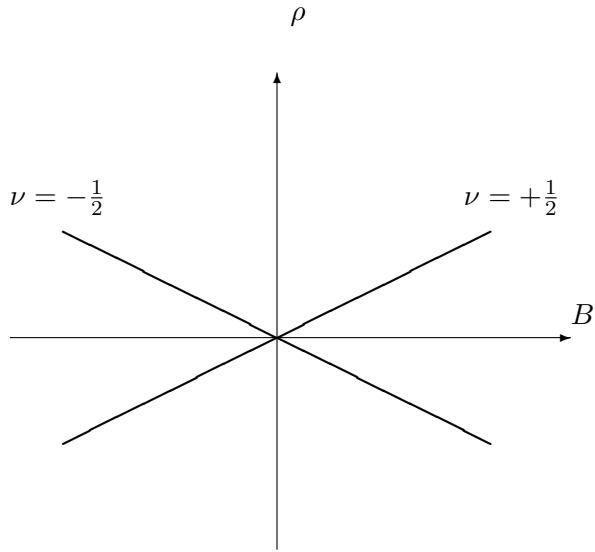
\begin{figure}
\begin{center}
\begin{picture}(240,240)(-20,10)
\put(0,100){\vector(1,0){210}}
\put(100,20){\vector(0,1){180}}
\thicklines
\put(20,60){\line(2,1){160}}
\put(20,140){\line(2,-1){160}}
\put(0,150){$\nu=-\frac{1}{2}$}
\put(170,150){$\nu=+\frac{1}{2}$}
\put(105,220,){$\rho$}
\put(210,105,){$B$}
\end{picture}
\end{center}
\caption{ {\bf The phase diagram of the D3-probe-D7 system}:  The defect has a charge gap only on the lines $\nu=\frac{1}{2}$
and $\nu=-\frac{1}{2}$. Elsewhere in the diagram it has metallic or semi-metallic behaviour.  The chiral condensate 
$\left<\bar\psi\psi\right>$ is nonzero at the origin and in the entire rest of the diagram except on the vertical and horizontal axes 
(away from the origin) where it vanishes. States
(away from the origin) on the horizontal axis preserve CP symmetry whereas states on the vertical axis (excluding
the origin) preserve $P$.  Both of these forbid a chiral condensate which is odd under $CP$ and $P$. (These symmetries
could be broken spontaneously.  Our numerical investigations of the axes away from the origin have found no symmetry breaking
states.)
At the origin, $C$ is preserved whereas $P$ and $CP$ are broken by the chiral condensate.  
 The four quadrants of the phase diagram are related by symmetries. $P$ corresponds to a reflection  through the vertical axis and $CP$ reflects
 through the horizontal axis. 
   \label{fig1}}
\end{figure}

We will  review the argument that the D3-D7 system is the string 
theory dual of a defect quantum field theory where the defect is an infinite, flat, 2+1-dimensional
space-time bisecting the 3+1-dimensional bulk.  The degrees of
freedom living on the defect are 2+1-dimensional Dirac fermions which interact by exchanging the quanta
of ${\mathcal N}=4$ supersymmetric Yang-Mills theory which occupies the surrounding 3+1-dimensional bulk.   
We will study the ground state of this system at charge neutrality, as well as the system with a constant charge density
and a constant external magnetic field.  As usual, the semiclassical string theory, with which we do computations,  solves the field 
theory in the large $N$ 't Hooft limit and the subsequent strong coupling limit.  Stability issues in both the string and field theory
force us to use an explicit finite ultraviolet cutoff at the strong coupling limit, so that all of the results what we quote below apply
to the field theory with a large but finite cutoff, far away from the scaling limit were the cutoff could be removed. 
We shall introduce an explicit cutoff by embedding the D7-brane in the full black D3-brane metric. (See the discussion after
equation (\ref{sd7}).)

Our  results are summarized in the phase diagram which is shown in figure \ref{fig1}.
It examines states of the system at zero temperature and with a constant charge density, $\rho$, and a constant
external magnetic field, $B$.  A third dimensional parameter is the ultraviolet cutoff and, typically, our numerical results 
concentrated on the region of the phase diagram where the cut-off
  is a factor of ten larger than either $\sqrt{B}$ or $\sqrt{\rho}$.  
To within the accuracy of our numerical computations, we find that the solution of 
the strong coupling theory is unique for any values of $B$ and $\rho$ except at the origin
$(B,\rho)=(0,0)$ where there are a constant symmetry preserving solution as well as two degenerate  and energetically favored, symmetry breaking solutions.  
The latter solutions spontaneously break the
parity, $P$,  and $CP$, where $C$ is charge conjugation symmetry of the theory at that point. 
The field theory order parameter for these
symmetries is the expectation value of 
a composite operator, the chiral condensate $<\bar\psi\psi>$ which transforms by a sign flip under both $P$ and $CP$. 
The string theory signal of this condensate is the pinching off of the D7 brane at a finite radius in the black D3 brane
geometry. 

The two  symmetry breaking
solutions at $(B,\rho)=(0,0)$   
are strongly coupled fractional topological insulators.  They describe two phases which each have a charge gap and they each 
have half of a unit of integer Hall conductivity per colour, $\sigma_{xy}=\pm\frac{N}{2}\cdot \frac{e^2}{2\pi}$ (even though both the
magnetic field and the charge density are zero).
If we choose one of these solutions, say the one with $\sigma_{xy}=+\frac{N}{2}\cdot \frac{e^2}{2\pi}$,
the charge gap survives as we move away from the origin along the line with Landau level filling fraction $\nu=\frac{1}{2}$ in figure \ref{fig1}, the
chiral condensate remains non-zero and the Hall conductivity remains at $\sigma_{xy}=+\frac{N}{2}\cdot \frac{e^2}{2\pi}$ along the entire line.
Similarly, along the $\nu=- \frac{1}{2}$ line, the
Hall conductivity is $\sigma_{xy}= - \frac{N}{2}\cdot \frac{e^2}{2\pi}$ and it 
connects to the $\sigma_{xy}= - \frac{N}{2}\cdot \frac{e^2}{2\pi}$ state
at the point $(B,\rho)=(0,0)$.  \footnote{ A mechanism for forming Hall plateaus is absent here in the ``clean'' system what we
are discussing here.  The incompressible Hall states
therefore appear only at specific filling fractions $\nu=\pm\frac{1}{2}$, rather than finite intervals of $\nu$.}

The only charge gapped states that we are able to find occur along the lines $\nu=\pm\frac{1}{2}$.  Away from 
these lines, the charge gap always vanishes.  At generic points in the phase diagram, away from
$(B,\rho)=(0,0)$ and away from the $B=0$ or $\rho=0$ axes, the discrete symmetries parity, $P$, charge conjugation, $C$,
and $CP$ are broken explicitly, as both of the parameters $\rho$ and $B$ are nonzero.  The chiral condensate $<\bar\psi\psi>$
  is not protected
by symmetry in these regions and it is generically nonzero.   On the $B=0$
 and the $\rho=0$ axes the system has either $P$ or $CP$ symmetry, respectively.  
  On the  lines $B=0$
 and $\rho=0$ , to the accuracy
 which we can determine, the solution of the theory does not break the respective $P$ and $CP$ symmetries
 and the chiral condensate $<\bar\psi\psi>$ vanishes there.  The axes, away from $(B,\rho)=(0,0)$ are the 
 only place in the phase diagram where $<\bar\psi\psi>=0$. 
 
 This leaves a discontinuity as
 the origin is approached along either of the axes, where a unique solution with vanishing 
 condensate suddenly becomes
 two degenerate, symmetry breaking solutions with non-zero condensates when the axes reach the origin.  
 
 Similarly, when the origin is approached along any other line (apart from the lines $\nu=\pm \frac{1}{2}$), 
 a unique ungapped and non-constant solution gets replaced by two degenerate gapped solutions.
   The details of the phase diagram  in the immediate vicinity of 
 the origin clearly requires a more detailed study than we have been able to do thus far. 
  We shall leave this for a future investigation. 
 
Our results contain a confirmation of the no-renormalization theorem for the Chern-Simons term for a U(1) gauge field. 
The Coleman-Hill theorem  \cite{Coleman:1985zi,Semenoff:1988ep}
shows that, in a 2+1-dimensional gauge field theory with U(1) gauge symmetry, 
the coefficient of the Chern-Simons term does 
not renormalize beyond the one-loop order in a charge-gapped system. This theorem has a perturbative proof which
holds order by order in perturbation theory in the coupling constant. 
Our work confirms that,  as long as the system has a mass gap,  that is, along the $\nu=\pm\frac{1}{2}$ lines in figure \ref{fig1}, 
the planar strong coupling limit indeed matches
what is expected from perturbation theory, and thereby demonstrates no-renormalization in a regime
where perturbation theory is not valid, at least for the system what we consider. 

Another observation that we shall make is that, despite having a single  species 
of fermions in 2+1-dimensions, coupled to a gauge field  and 
transforming in the fundamental representation of a non-abelian gauge symmetry, 
the parity anomaly is absent in the defect field theory. 
When the system has a charge gap, integrating out the fermions would induce a non-Abelian Chern-Simons term for the bulk fields 
integrated over the defect and with a half-quantized coefficient \cite{Niemi:1983rq}-\cite{Redlich:1983kn}.  
If it were an isolated 2+1-dimensional quantum field theory, 
gauge invariance would require that the coefficient of the Chern-Simons term is  quantized.  The half-quantization would lead to the parity
anomaly.    However, in the defect theory, we shall
argue that the Chern-Simons term has a gauge invariant extension to the bulk, see equation (\ref{defectcs}),  
and the gauge symmetry remains intact.   

There have been a number of works which seek holographic systems which exhibit both integer and fractional quantum Hall effect 
\cite{Krishnan:2009nw}-\cite{Fujita:2009kw}
 The present
work is closest in spirit to that in references \cite{Bergman:2010gm}-\cite{Kristjansen:2013hma} in that the incompressible state that we find is made possible by the presence of the Wess-Zumino term in the action for the D7 brane.

 In the next two subsections, we shall review the properties of the D3-D7 system in string theory and the defect gauge field theory which 
 is dual to it, respectively.    In Section 2 we shall describe the geometrical problem of solving the semi-classical limit of the string theory
 in some detail.  In Section 3, we discuss general properties of the equations which determine the D7 brane geometry.  In Section 4, we summarize
 our numerical solutions of these equations. In Section 5, we present further discussion of our results.

\subsection{The D3-D7 string theory system}

 The D3-probe-D7 brane system, oriented so as to have 2+1-dimensional Poincare symmetry, 
 is an interesting example of probe brane holography. In flat space, the relative orientations of the
 D3 and D7 branes are given by 
 \begin{align}\label{table}
\begin{array}{ccccccccccc} ~~ & 0 & 1 & 2 & 3 & 4 & 5 & 6 & 7 & 8 & 9 \cr
                D3 & X & X & X & X & O & O & O & O & O & O \cr
                D7 & X & X & X & O & X & X & X & X & X & O \cr \end{array}
\end{align}
where the branes extend in the directions $X$ and sit at points in the directions $O$. 
This is an $\#ND=6$ brane intersection \cite{Polchinski:1998rr}.  It is not supersymmetric.  However, it is tachyon free,
indicating that, in the limit of weak string coupling, it is stable. 
 The only low energy modes of the open strings which connect the D7 and D3 branes are in
 the Ramond sector and they correspond to defect fermions which are two-component spinors of the SO(2,1) Lorentz
 group.  
  For this reason, the D3-D7 system 
 has been used to formulate holographic models  of relativistic
 two dimensional electron gases \cite{Rey:2008zz}-\cite{Grignani:2016npu}.

  The D3 and D7 branes can be separated along the 9-direction in (\ref{table}).  This separation corresponds to 
  a mass for the defect
  fermions.  As we will discuss shortly, the separation breaks a discrete symmetry of the theory which corresponds to parity in
  the defect quantum field theory. It can either be input to the
theory, which corresponds to adding an explicit mass term for the fermions
  in the dual gauge theory, or, as we shall see in some cases, it can happen spontaneously, and the theory then has spontaneously
  broken parity and time reversal symmetry and the gauge theory has a chiral condenate $<\bar\psi\psi>\neq 0$. 
   
  In the large N, 't Hooft limit, and in the strong coupling limit, the D3 branes are replaced by the  $AdS_5\times S^5$ background and 
  the mathematical problem of solving the theory is to find the embedding of the D7-brane in that background
  geometry, subject to the appropriate boundary conditions.  
  The most symmetric embedding of the D7 brane in  $AdS_5\times S^5$  would have a worldvolume with the geometry $AdS_4\times S^4$.  However,
 any embedding of a D7 brane which is asymptotically $AdS_4\times S^4$ turns out to be unstable.  The equation for the fluctuations of the position of the $S^4\subset S^5$ contains
 a tachyonic mass term which violates the Breitenlohner-Freedman 
  bound for excitations on  $AdS_4\times S^4$.  This will pose a difficulty for using this brane configuration for holography.  
  
 Recall that  the D3-D7 system was stable in the limit of weak string coupling where the string spectrum was
 free of tachyons. 
 The appearance of an instability in the strong coupling limit must therefore be due to strong coupling dynamics, 
 and of course, the presence of an instability implies that a tachyon must appear
 in the spectrum.  In this case, the tachyon is a bound state of two strings, a fluctuation of the position of the $S^4$ which the D7 brane
 worldvolume wraps in $S^5$.  This mirrors the fact that in the field theory dual the instability
 is the signal of a phase transition for which the order parameter is a composite operator.

The mathematical problem of solving the semiclassical string theory is to determine the embedding of the D7 brane in the background string fields which are sourced by a stack of $N$ D3 branes which, in the appropriate low energy limit, is $AdS_5\times S^5$.    This involves extremizing the
  the D7 brane action which is a combination of the Born-Infeld action and the Wess-Zumino terms,
\begin{align}
S_{D7}=-T_7\int  \sqrt{\det( g+2\pi\alpha' F)}+
\frac{(2\pi\alpha')^2T_7}{2}~\int F\wedge F\wedge C^{(4)},
\label{sd7}
\end{align}
where $g$ is the metric of the D7 brane worldvolume (extracted from (\ref{metricd7}) below) and $F$ is the field strength of a world-volume
U(1) gauge field. 
   However, because of the instability which appears when we attempt this, 
in order to find a stable embedding, we shall embed the D7 brane in the full black D3 brane metric, rather than $AdS_5\times S^5$ which is
 its near-horizon geometry.   This approach was  suggested by Davis, Kraus and Shah \cite{Davis:2008nv}. 
 The D7 brane has a stable embedding  in the black D3 brane geometry which has the appropriate symmetries to match the embedding that we
 would expect to find in $AdS_5\times S^5$.  As a regularization of the quantum field theory, 
 we can view this as a stringy resolution of the ultraviolet by 
 undoing the scaling limit which produced AdS/CFT in the first place.   The fundamental string tension $\alpha'$ will now appear in physical quantities  and
 it will play the role of an ultraviolet cutoff.  
The strategy will be to replace the metric of $AdS_5\times S^5$,  \footnote{
The Ramond-Ramond 4-form in these coordinates is  
\begin{align}\label{4form}
C^{(4)}=  \lambda{\alpha'}^2\left[ 
r^4dt\wedge dx\wedge dy\wedge dz+4  c(\psi)  d\Omega_4\right],
\end{align}
where
$
\partial_\psi c(\psi)=\sin^4\psi
$
and  
$
c(\psi)=   -\frac{1}{4}\sin^3\psi\cos\psi-\frac{3}{8}\cos\psi\sin\psi+\frac{3}{8}[\psi-\pi/2] 
$. Note that we have chosen an integration constant so that $c(\pi-\psi)=-c(\psi)$.  This differs from the choice in ref. \cite{Mezzalira:2015vzn} where
the constant was fixed by requiring that $ c(\psi) d \Omega_4 $ is non-singular when the $4$-sphere shrinks to a point.  It is easy to
see that, with proper care, physical quantities do not depend on this choice.}  
\begin{align}
ds^2=R^2\left[ r^2  \left(- dt^2+dx^2+dy^2+dz^2\right) 
+  \frac{dr^2}{ r^2}
\right. 
\left.+d\psi^2+\sin^2\psi d\,\Omega_4^2 \right],
\label{metricads}\end{align}
by the black D3-brane metric,
 \begin{align}
ds^2=R^2\left[\frac{r^2}{\sqrt{f(r)}} \left(- dt^2+dx^2+dy^2+dz^2\right) 
+ \sqrt{f(r)}\left( \frac{dr^2}{ r^2}+d\psi^2+\sin^2\psi d\,\Omega_4^2\right)\right],
\label{metricd3}\end{align}
where $d\Omega_4^2$ is the metric of $S^4$ and $\psi\in[0,\pi]$,
$
f(r)=1+R^4r^4 
$ and  
$
R^2=\sqrt{\lambda}{\alpha' } 
$.\footnote{By further rescaling $r\to  r/R\Lambda$ and $x^\mu\to R\Lambda x^\mu$ we could replace
the cutoff $\frac{1}{R}$ by $\Lambda$.}
Here, the radial coordinate $r$ has been scaled by 
powers of $R$ so that it has
the units of energy, whereas $R$ itself has units of distance.  Aside from the overall factor of $R^2$, the metric still depends
on the fundamental string tension $\alpha'$ through the 
dependence of $f(r)$ on $R$.  In this dependence, $\frac{1}{R}$ has the dimension of energy and it plays
the role of  the ultraviolet cut-off.  
When $r<<1/R$, and $\sqrt{f(r)}\sim 1$, the geometry is $AdS_5\times S^5$.
When $r>>\frac{1}{R}$ and $\sqrt{f}\sim r^2R^2$, it is  ten dimensional Minkowski space.  Now, since the boundary of the spacetime
and the boundary of the D7 brane described by the embedding (\ref{metricd7}) are not in $AdS$ geometry
, we do not expect that the
tenets of holography hold. (The bulk geometry does not have any scaling symmetry.  It is not conformally AdS  or Lifshitz or one of the other 
generalizations of AdS which are used for holography.)
We shall nevertheless be able to extract some features of the field theory.   Similar computations have
been carried out in holographic constructions of double-monolayers of Weyl semimetals  \cite{Grignani:2014tfa,Grignani:2016npu}. 

The embedding of the  D7 brane in the spacetime (\ref{metricd3}) is mostly dictated by symmetry.   We shall
embed it so that it wraps the coordinates $t,x,y,r,\Omega_4$ and it sits at
a point in the coordinates $z,\psi$.  What is more, its $\psi$-coordinate can depend on $r$, so that the worldvolume metric
is 
 \begin{align}
d\sigma^2=R^2\left[ \frac{r^2}{\sqrt{f(r)}}  \left(- dt^2+dx^2+dy^2\right) 
+ \sqrt{f(r)} \left(\frac{dr^2}{ r^2}(1+r^2\psi'(r)^2)+\sin^2\psi d\,\Omega_4^2\right)\right],
\label{metricd7}\end{align}
Parity symmetry is the transformation $(x,\psi)\to (-x, \pi -\psi)$ and parity symmetric solutions must have 
$\psi=\frac{\pi}{2}$.\footnote{ The Wess-Zumino action of the D7-brane, $\int F\wedge F\wedge C^{(4)}$, where $F$ are field
strengths of worldvolume gauge fields, is also symmetric under this parity transformation.}

The most symmetric embedding has constant $\psi=\frac{\pi}{2}$.  It approaches eight-dimensional Minkowski
space when $r>>1/R$ and $AdS_4\times S^4$ when $r<<1/R$.  The large $r$ region is now stable, however, the small $r$
region is still  $AdS_4\times S^4$ and it is still unstable to fluctuations of the embedding.     
There is a more stable solution where $\psi$ becomes $r$-dependent.  This solution breaks parity symmetry. 
 Once $\psi$  is $r$-dependent,  there are two solutions with the same free energy, with either $0\leq \psi(r)\leq\frac{\pi}{2}$ or $\pi\geq \psi(r)\geq\frac{\pi}{2}$,
 and which are related by parity.  For a particular set of solutions, the so-called gapped solutions, 
the radius of the $S^4$ decreases as $r$ decreases from infinity, until it goes to zero at a finite radius, $r=r_0$. 
This shrinking cycle allows the brane geometry to end smoothly at   $r_0$. 

This embedding of the $D7$-brane has the electromagnetic response of an insulator.  In the string theory, 
the defect fermions, and U(1) charge carriers, are strings which
stretch from the D3 brane to the D7 brane, here from the Poincare horizon at $r=0$ to the D7 brane's closest approach at $r=r_0$.  
These strings now have a minimum
length and a non-zero string tension.  Thus, they have a mass gap.  This gap is proportional to the only dimensional parameter in the problem,
the ultraviolet cutoff $\frac{1}{R}$.  In this solution, we see a dynamical reason why the D7 brane embedding in $AdS$ was unstable.  The D7 brane
simply does not come to the near horizon region of the D3 branes.  The stable embedding truncates at a radius of order $\frac{1}{R}$, rather than
in the region $r<<\frac{1}{R}$ which would be needed.  In principle, we could attempt to include corrections in $\frac{1}{\sqrt{\lambda}}$  by including
fluctuations of the string worldsheet sigma model. The radius $r_0$ where the D7 brane ends would then become a function of 
$\lambda$ and, if the hypothesis that
the chiral phase transition exists is correct and if it is a continuous phase transition, 
we could adjust $\lambda$ to be near $\lambda_c$ 
so that the D7 brane approaches the Poincare horizon and take a limit where it occupies the near horizon region,
corresponding to removing the cutoff in the field theory.      This, of course, is identical to the procedure of removing the cutoff
by tuning the coupling constant to the fixed point in the quantum field theory.

\subsection{The defect quantum field theory}

 The quantum field theory dual of this D3-D7 system is a defect field theory with
 fermions occupying an infinite planar 2+1-dimensional defect that is embedded in 3+1-dimensional
 spacetime.  The 3+1-dimensional bulk contains 
 ${\mathcal N}=4$ supersymmetric Yang-Mills theory and the defect fermions interact by exchanging quanta
 of the ${\mathcal N}=4$ theory.   A Lagrangian for this defect field theory is   
 \begin{align}
  S=\int d^4x  \frac{1}{g_{\rm YM}^2}{\rm Tr}  \left\{ -\frac{1}{2 } F_{\mu\nu}F^{\mu\nu}+D_\mu\Phi^a D^\mu\Phi^a+\ldots\right\}
  +\int d^3x   \bar\psi(i\slashed\partial +\slashed A   +\Phi^6)\psi,  
   \label{action} \end{align}
   where the first, four-dimensional integral is the action of ${\mathcal N}=4$ Yang-Mills theory and the second, three-dimensional
   integral is over the defect world-volume and it contains the defect fermions. 
 The defect breaks the R symmetry of the ${\mathcal N}=4$ theory from SO(6) to SO(5) by coupling to one of the scalar fields. The
 defect fermions
 are SO(5) singlets and they transform in the fundamental representation of the $SU(N)$ gauge group.  They also have a global $U(1)$ symmetry. 
We will use the conserved charge for this symmetry to model electric charge, analogous to the electric charge of a layer of material such as
the electrons in 
a monolayer of graphene.  We will deform the theory by introducing a density  
 for this charge, as well as coupling it to an external magnetic field. Our goal will be to explore the electromagnetic properties of the defect.
Our work will be at zero temperature, but could easily be extended to finite temperature.  The string theory computations which we
do examine the field theory (\ref{action}) in the planar, large $N$ 't Hooft limit, where $N\to\infty$ keeping $\lambda\equiv g_{\rm YM}^2N$ finite
and the subsequent large $\lambda$, strong coupling limit.

   A fermion mass term in 2+1-dimensions violates parity and time reversal invariance.   As well as the fermion, parity transforms
  $\Phi^6(x)\to -\Phi^6(x')$, where $x'$ has a reflection of one of the spatial coordinates which lie inside the defect. 
  The corresponding transformation in string theory is  a  rotation by $\frac{\pi}{2}$ radians in a plane of 
  ten dimensional flat space, and a similar discrete symmetry of $AdS_5\times S^5$. Both are symmetries of the IIB string theory
  in those backgrounds.  If parity and time reversal invariance were absent, we would need to add marginal operators constructed from the bulk
  ${\mathcal N}=4$ fields to the defect action
  \begin{align}\label{defectaction}
  \frac{\kappa}{4\pi} \int d^3x~ {\rm Tr}( AdA+\tfrac{2}{3}A^3)~,~
  \int d^3x~ (\Phi^6)^3~,~
\int d^3x~ \bar\chi\Gamma^9\chi~, \end{align}
where $\chi$ is the ${\cal N}=4$ fermion.
  All of these operators are odd under parity and would be absent from the action of  the parity symmetric theory.  However, if, as we shall find, parity is spontaneously
  broken, these operators must be added to the defect action as the first two would be induced by integrating out the worldsheet fermions whereas
  the last one is allowed by symmetry but could only appear from higher order corrections. 
   Renormalization would be expected to give some of the  coefficients, as well as the
  Yukawa coupling to worldsheet fermions non-trivial beta-functions.  A better understanding of the renormalization of this model would be
  interesting but is beyond the scope of this paper.  

As we have reviewed, the D3-D7 system is tachyon-free at weak coupling and it becomes unstable
at strong coupling. 
On the field theory side, this instability  is  interpreted as an instability of the defect
 field theory to generating a parity and time reversal-violating condensate,   $\left<\bar\psi\psi\right>$ 
 when the coupling constant $\lambda$ is large enough \cite{Kutasov:2011fr}.
 At small values of the ${\mathcal N}=4$ coupling, $\lambda$, it is expected that parity is a good symmetry.  
 As  $\lambda$ is increased from small values, there is a critical coupling, $\lambda_c$ where
 there is a phase transition.  In the strong coupling phase, when $\lambda\geq \lambda_c$,  
 a condensate  forms 
  and parity symmetry is spontaneously broken.  The   form of the condensate suggested by approximate solutions of Schwinger-Dyson equations  and a 
  renormalization group analysis is $<\bar\psi\psi>\sim \Lambda^2 e^{-1/\sqrt{\lambda-\lambda_c} }$ \cite{miransky,Kaplan:2009kr}.
Assuming that there is a second order quantum phase transition at $\lambda=\lambda_c$, we then know that 
the chiral condensate can become cutoff independent only when $\lambda$ is lowered and finely tuned to $\lambda_c$. 
 The limit of the field theory that we are solving by examining the semi-classical string theory, on the other hand, 
 is where  $\lambda$ is very large, putatively well away from
 the point $\lambda_c$ where the cutoff can be removed.

 This fact raises a fundamental difficulty for doing holography with the D3-D7 system in this configuration.                                  Even in the
 quantum field theory the coupling constant can be tuned only once the theory has an ultraviolet regularization and, it usually becomes regularization-independent
 only at certain values of the coupling which are fixed points of the renormalization group flow. 
  If we set the coupling constant to other than a fixed point value, we expect   non-universal regulator-dependent behaviour 
   and the field theory description is of limited utility.   In the following, we will
argue that we can nevertheless extract  a few essential   features of the theory by studying its holographic dual with an ultraviolet cutoff. One is  an answer to 
the generic question as to whether the defect has a charge gap.
Since parity forbids a fermion mass, we would expect that parity invariant solutions of the theory are gapless, whereas parity broken phases have a 
 charge gap.  If we can regulate the theory using a parity invariant regularization, we can ask whether parity is broken or equivalently whether the
theory has a gap.   

 The dual quantum field theory  has   action (\ref{action}). The defect fermions have an SU(N) gauge invariant coupling 
to the ${\cal N}=4$ vector potential.  If they also become massive due to parity symmetry breaking, integrating them out generates
an effective action for the defect in the form of equation (\ref{defectaction}) where, if $m$ is the mass, the Chern-Simons level is
$\kappa={\rm sign}(m)/2$ and the other coupling constants contain ${\rm sign}(m)$.   If this were a stand-alone three-dimensional
spacetime, the Chern-Simons term with a half-integer quantized level would not be gauge invariant when $N$ is odd and this theory 
would have an anomaly.  However, in the defect theory, the large gauge transformations which transform the Chern-Simons term would
have to be extendable to the bulk, and they are not.  An easy way to see this is to note that the Chern-Simons term can be written as
a gauge invariant integral over the bulk, 
 \begin{align}
   \frac{{\rm sign}(m)}{4\pi} \int d^3x {\rm Tr} ( AdA+\frac{2}{3}A^3) =-\frac{1}{8\pi}\int d^4x{\rm sign}(m(z-z_0)) {\rm Tr} F\wedge F.
\label{defectcs}  \end{align}
  The latter is a space-dependent theta-term for the bulk ${\mathcal N}=4$  Yang-Mills theory. Unlike the other terms in (\ref{defectaction}), the Chern-Simons term is not expected to  renormalize.  If the other defect operators get positive anomalous dimensions, they will be suppressed by powers of the cutoff and 
  the trace of the defect that remains is a jump in the theta-angle at the defect's location.
  
  The defect also has an interesting electromagnetic signature.   If we couple a non-dynamical U(1) vector potential $a_\mu(x)$ to the defect
  fermions by replacing $\partial_\mu$ with $\partial_\mu+iea_\mu$, when the fermions are integrated out, the effective action for $a_\mu(x)$ will be, in the leading orders in a derivative expansion, the abelian
  Chern-Simons term,
  \begin{align}
  S_{\rm eff} = ~\frac{Ne^2{\rm sign}(m)}{8\pi}~\int d^3x~ ada + {\mathcal O}(1/m).
  \label{emresponse}
  \end{align}
  This result is obtained from a one loop diagrammatic computation \cite{Niemi:1983rq}-\cite{Redlich:1983kn}
and, as long as the theory retains a charge gap, it does not obtain corrections from higher
  orders in perturbation theory  \cite{Coleman:1985zi,Semenoff:1988ep}. 
 Functional derivatives give the current-current correlation function, 
 at small momenta,
 \begin{align}\label{current-current}
\left< j_\mu(p) j_\nu(-p)\right> = \frac{Ne^2}{4\pi}{\rm sign}(m) \epsilon_{\mu\nu\lambda}p^\lambda +{\mathcal O}(p^2),
\end{align}
from which we can deduce that the defect is an insulator
\begin{align}
\sigma_{xx}=0,
\end{align}
and that it has a Hall conductivity
\begin{align}\label{zerofieldhall}
\sigma_{xy} ~=~ \frac{Ne^2}{4\pi}~{\rm sign}(m)~=~\frac{N}{2}\frac{e^2}{2\pi}~{\rm sign}(m) .
\end{align}
This half-quantized (for each of the N colours) ``zero field''  Hall  conductivity appears in the 
absence of an external magnetic field and in the absence of a charge
density. In quantum field theory, there is a no-renormalization theorem for the abelian Chern-Simons
term.   As long as the theory retains a charge gap, all higher order contributions cancel, order by 
order in perturbation theory.  It would be interesting to know
whether this cancelation occurs when the theory is strongly coupled, beyond the radius of
convergence of perturbation theory.   Remarkably, we shall be able to give an affirmative answer to this question
using holography.

In a very interesting paper,  Hoyos, Jensen and Karch \cite{HoyosBadajoz:2010ac} discuss a  holographic model of
a fractional topological insulator which is also based on a D3-D7 brane system but with the D7 brane oriented so that
it has 3+1-dimensional Poincare invariance.  In that supersymmetric configuration, the low energy degrees of freedom 
of the D3-D7 strings are a 3+1-dimensional hypermultiplet which transforms in the fundamental representation of the SU(N) gauge group.
They then give the hypermultiplet a mass which changes sign on a 2+1-dimensional defect and they then use an axial
gauge transformation and the axial anomaly to argue that the effective action contains the term on the right-hand-side of
equation (\ref{defectcs}).  Indeed, in the limit where the hypermultiplet mass is very large, their construction should be
very similar to our model.   The 3+1-dimensional Dirac fermions with such a space-dependent mass term should have 
Jackiw-Rebbi zero modes \cite{Jackiw:1975fn}
which propagate like massless fermions on the defect and they would have the properties similar to 
our defect fermions, the main difference being that they preserve SO(4) rather than SO(5) R-symmetry. Scalar fields would not
have these zero modes, so supersymmetry would be broken.  In the large mass limit, these zero modes would not decouple 
and they could be seen in the low energy states of the defect.
Then, in the strong coupling limit, strong coupling dynamics could break the parity symmetry of the defect theory 
and lead to the defect Chern-Simons effective action in (\ref{defectcs}). One difference with us is their assumption that 
quarks are confined so that the basic unit of charge is $\tilde e =eN$, which would make the Hall conductivity
$\sigma_{xy}=~
\frac{1}{2N}~ \frac{\tilde e^2}{2\pi}
$.  Whether the defect fields would be confined is an interesting question. Some   implications of 
the effective field theory, ${\mathcal N}=4$ Yang-Mills theory with a theta-angle which jumps at the defect, 
have been discussed  by Estes et. al. \cite{Estes:2012nx}.

  \section{The geometric set-up}

The metric of extremal black D3 branes is given in (\ref{metricd3}), copied below,
$$
ds^2=R^2
\left[ f^{-1/2}(r)r^2 \left(- dt^2+dx^2+dy^2+dz^2\right) +
\right. 
\left. 
+
f^{1/2}(r)\left( \frac{dr^2}{ r^2}+d\psi^2+\sin^2\psi d\,\Omega_4^2\right)\right],
$$
 In these coordinates, the  $S^5$ is written as a fibration   by 
4-spheres over the interval $\psi\in[0,\pi]$.  
Furthermore,
$(t,x,y,z,r)$ are coordinates of the Poincare patch of the near horizon geometry, $AdS_5$,  and
$$
f(r)=1+R^4r^4.
$$
The square of the radius of curvature is 
$$
{\lambda}{\alpha' }^2= R^4. 
$$
The Ramond-Ramond 4-form of the IIB supergravity background takes the form
\begin{align}\label{4form}
C^{(4)}=  \lambda{\alpha'}^2\left[ 
r^4dt\wedge dx\wedge dy\wedge dz+4  c(\psi)  d\Omega_4\right].
\end{align}
Here,
$
\partial_\psi c(\psi)=\sin^4\psi
$
which we integrate as
\begin{align}
c(\psi)=   -\frac{1}{4}\sin^3\psi\cos\psi-\frac{3}{8}\cos\psi\sin\psi+\frac{3}{8}[\psi-\pi/2]. 
\end{align}
The choice of integration constant is a string theory gauge choice and our
results will not depend on it.  We have chosen it so that $c(\pi/2)=0$. 

We will study the D7 brane  in the background (\ref{metricd3}) and (\ref{4form}).  
We will take the D7 brane world-volume as wrapping the $S^4$ (thus having $SO(5)$ symmetry)
and extending in the directions $r,x,y,t$.  The brane will sit at a point in the $z$-direction and at an $r$-dependent point
in the $\psi$-direction.  The induced metric is given in equation (\ref{metricd7}) and the mathematical problem 
reduces to finding extrema of the Born-Infeld action, given in equation (\ref{sd7}). The world-volume metric, $g_{ab}$,
in that equation is the one that is extracted from the ansatz (\ref{metricd3}).  The detailed form of the Wess-Zumino term
with this ansatz will be discussed in the next section. 


\subsection{The Wess-Zumino term}

The reasoning for the normalization of $C^{(4)}$  
follows Davis et.\ al.\ \cite{Davis:2008nv} who argue that it should obey a condition which fixes
the flux through the five-sphere, 
\begin{align}
\int _{S^5}dC^{(4)}= 2\kappa_{10}^2 T_3N,
\end{align}
where $T_3$ is the  D3 brane tension and $\kappa_{10}$ is the ten-dimensional supergravity Newton constant
which, in terms of the string coupling is
\begin{align}
2\kappa_{10}^2=g_s^2(2\pi)^3(2\pi\alpha')^4.
\end{align}
The D3 and D7 brane tensions are
\begin{align}
T_3=\frac{1}{g_s(2\pi)(2\pi\alpha')^2}~,~~T_7=\frac{1}{g_s(2\pi)^3(2\pi\alpha')^4}.
\end{align}
We begin with taking $dC^{(4)}$ to be proportional to the volume form of the unit five-sphere, 
\begin{align}
dC^{(4)}=\gamma \sin^4\psi d\psi\wedge d \Omega_4. ~ 
\end{align}
The volume of the unit $S^5$ is $\pi^3$ and of $S^4$ is $\frac{8}{3}\pi^2$. 
We need to determine the constant $\gamma$
\begin{align}
\int_{S^5}dC^{(4)}=\pi^3\gamma  = 2\kappa_{10}^2 T_3N =g_s^2(2\pi)^3(2\pi\alpha')^4\frac{1}{g_s(2\pi)(2\pi\alpha')^2}N.
\end{align}
We conclude that
\begin{align}\gamma =\frac{1}{\pi^3}g_sN (2\pi)^2(2\pi\alpha')^2,
\end{align}
and
\begin{align}
C^{(4)}= \frac{1}{\pi^3}g_sN (2\pi)^2(2\pi\alpha')^2 c(\psi)d\Omega_4,
\end{align}
which, using $\lambda =4\pi g_s N$,  we can re-organize  to get
\begin{align}
C^{(4)}= 4\lambda{\alpha'}^2c(\psi)d\Omega^4.
\end{align}
Now, the Wess-Zumino term is
\begin{align}
S_{WZ}=\frac{1}{2}(2\pi\alpha')^2T_7~\int F\wedge F\wedge C^{(4)}
=\frac{N}{8\pi}\int F\wedge F \cdot\frac{16}{3\pi}c(\psi),
\label{wz}\end{align}
where we have used the fact that the unit four-sphere has volume $8\pi^2/3$. In the final equation (\ref{wz}), 
we have normalized the factor with $c(\psi)$ so that, $\frac{16}{3\pi}c(\psi=0)=1$ and $\frac{16}{3\pi}c(\psi=\pi)=-1$.

\subsection{Equation of motion for the gauge fields\label{symmetries}}

The DBI plus WZ action for the probe D7-brane with world volume gauge field strength $F$ was given in eqn.~(\ref{sd7}). 
 From this we derive the following equation of motion for the gauge field
 $$
-\partial_a\left( \frac{d}{dF_{ab}} \left[- \sqrt{-\det(   g+2\pi\alpha' { F})} +\frac{(2\pi\alpha')^2}{2}
  {F}\wedge  { F}\wedge C^{(4)}\right]\right)=0.
  $$
 We will assume that the worldsheet metric and the gauge field strengths depend only on the coordinate $r$.
  We will also assume that the $S^4$ components of the gauge fields  vanish.   
  The Bianchi identity,  where we put the $\mu,\nu=x,y,t$, is
$$
\partial_\mu F_{\nu r}+ \partial_\nu F_{ r\mu}+ \partial_r F_{\mu\nu} =0,
$$
and it is satisfied by $x,y,t$-independent $F_{r\mu}$ only when $F_{\mu\nu}$ is 
independent of $r$. 
  If we introduce the displacement field
  $$
D^{ab}=-T_7\int_{S^4}~\frac{d}{d F_{ab}(\sigma) }\sqrt{-\det(g+2\pi\alpha' F)},
$$
the equation of motion for the gauge field reduces to 
 \begin{align}\label{displacement}
\frac{d}{dr}\left[ D^{r\nu}(r)+\frac{N}{4\pi}\frac{1}{2}\epsilon^{r\nu\lambda\rho}F_{\lambda\rho} \frac{16}{3\pi}c(\psi(r))\right]=0.
 \end{align}
The integral of this equation is 
\begin{align}\label{displacement1}
D^{r\nu}(r)+\frac{N}{4\pi}\frac{1}{2}\epsilon^{r\nu\lambda\rho}F_{\lambda\rho} \frac{16}{3\pi}c(\psi)=J^\nu,
\end{align}
where $J^\nu$ are constants of integration which are interpreted as the constant $U(1)$ charge and current density.
Since we will use boundary conditions where $c(\psi)$ vanishes at $r\to\infty$, $J^\mu$ is just
given by the asymptotic  
$J^\nu= D^{r\nu}(\infty)$. 

If the D7 brane has a charge-gapped (Minkowski) embedding, that is, one which pinches off at a finite radius, in order to be nonsingular, the radial electric field $D^{r\nu}$ 
must vanish at the pinch off radius. Moreover, the 4-sphere must shrink to a point, that is 
$\psi\to 0$ or $\psi\to \pi$. In that case $\frac{16}{3\pi}c(\psi\to0)\to 1$ or $\frac{16}{3\pi}c(\psi\to\pi)\to -1$.
These are degenerate solutions (they are related by symmetry).  Then, we must also have either
\begin{align}
\frac{N}{4\pi}~\frac{1}{2}\epsilon^{r\nu\lambda\rho}F_{\lambda\rho}& =J^\nu
\\
&{\rm or}   \nonumber
~~\\
\frac{N}{4\pi}~\frac{1}{2}\epsilon^{r\nu\lambda\rho}F_{\lambda\rho}& =-J^\nu,   
\label{csresponse}  
\end{align}
In particular, if there is a charge density and a magnetic field, these relationships are
$$
\frac{N}{4\pi}B =\pm\rho ~~,~~\frac{N}{4\pi} \epsilon_{ij}E_j  =\pm J_i. 
$$
The first of the above equations implies that the quantity which we would identify as 
$
\nu = \frac{2\pi\rho}{NB}.
$ would have to obey $
\nu=\pm \frac{1}{2},
$ for any charge gapped embedding of the D7 brane.   Conversely, the gapped embedding of a single
D7 brane with a constant magnetic field and charge density can only occur at the magic filling
fractions $\nu=\frac{1}{2}$ and $\nu=-\frac{1}{2}$.   
Moreover,the conductivity tensor is gotten from 
$$
J^i = \pm \frac{N}{4\pi}\epsilon^{ij}E^j~~,~~\sigma_{xy}=\pm\frac{N}{2}\frac{e^2}{2\pi}~~,~~\sigma_{xx}=0,
$$
which is valid whether there is a magnetic field or not.  Since it can occur in vanishing magnetic field, it is called
a ``zero field Hall effect''.
The states with $\nu=\pm\frac{1}{2}$ are fractional topological insulators, with vanishing longitudinal conductivity and half-integer quantized Hall conductivity. 

Finally, we note that there is no solution of the equation (\ref{displacement1}) where only two of the quantities $b$, $\rho$ and $\frac{\pi}{2}-\psi(r)$ (or, equivalently,
$\left<\bar\psi\psi\right>$)
are non-zero.  Any two of them being non-zero induces the third.  This is a result of the fact that they violate two symmetries, according to 
the following table (where $P$ is parity $C$ is particle-hole symmetry and $CP$ is their composite; yes means symmetric, no means not symmetric)
$$
\begin{matrix} ~ & P & CP & C \cr
   B& no & yes & no \cr
   \rho & yes & no & no \cr
  \left<\bar\psi\psi\right>  & no & no & yes\cr
   \end{matrix}
   $$
   Any two of these quantities being non-zero violates all of the symmetries which suppress the third one. In the following section,
   we will find states which are apparently stable and where only one of these objects is turned on.  These are states of higher symmetry - they preserve one of
   $P$, $C$ or $CP$.  In the case of $B$ and $\rho$, they appear on the axes of the phase diagram in figure \ref{fig1}. The case of $\left<\bar\psi\psi\right>$ nonzero with $B$ and $\rho$ both zero is at the origin of that diagram.  (Equivalent to the existence of a chiral condensate is the deviation of $\frac{\pi}{2}-\psi(r)$  from zero. ) 
     We also find that there is no stable state where none of the three $B,~
   \rho,~
  \left<\bar\psi\psi\right> $  are turned on.

\section{Equations of motion for the geometry}

We will study  D7 branes  in the background (\ref{metricd3}) and (\ref{4form}) by solving the equations of motion 
that are derived from the action (\ref{sd7}).  We will begin with the case of vanishing charge density and magnetic field.
Then we will add   charge density and magnetic field in the 
following subsection.

\subsection{No gauge fields\label{nogaugefields}}

  To begin, we note that the equations of motion for the world volume gauge fields will always be satisfied
  by putting those fields to zero.  Let us begin by considering this solution.   The 
  Dirac-Born-Infeld action then takes the form
  $$
  S_{7}= -{\cal N}_7 \int_0^\infty dr \, r^2f^{\frac{1}{2}}\sin^4\psi(r)\sqrt{1+(r\psi'(r))^2},
  $$
  where
  \begin{equation}
  \label{Snorm}
  {\cal N}_7= \frac{4}{3}\frac{\lambda N}{(2\pi)^4} V_{2+1},
  \end{equation}
  where $V_{2+1}$ results from the integration over $t,x, y$.
  The equation of motion for $\psi$ is
 
  \begin{align}
  \label{eomnogauge}
  \frac{-2}{1+e^{4y}R^4}\dot\psi + 5\dot\psi+\frac{\ddot\psi} { {1+\dot\psi^2}}
  = {4\cot\psi }, 
 \end{align}
  where $y=\ln r$ and $\dot\psi=\frac{d}{dy}\psi$.
  
  This equation permits a {\it maximally symmetric constant solution $\psi=\frac{\pi}{2}$}.  We shall 
  now give an argument to show that this solution is {\it \bf unstable}.   
  Upon linearizing the equation of motion about this solution, by setting $\psi=\frac{\pi}{2}-\delta\psi$, one gets
  $$
  \frac{-2}{1+e^{4y}R^4}\delta\dot\psi + 5\delta\dot\psi+\delta\ddot\psi 
  = -4\delta\psi. 
  $$
  At $r\to \infty$, it is solved by $\psi\sim r^\alpha=e^{\alpha y}$ with 
  \begin{align}\label{char1}
  \alpha^2+5\alpha+4=0~~,~~\alpha=-1,-4,
  \end{align}
  which furnish a satisfactory non-normalizable and normalizable mode at the $r\to\infty$ boundary of the worldsheet
  which is the boundary of eight-dimensional Minkowski space.   
  However, at $r\to 0$, where the geometry is $AdS_4\times S^4$,  it is  
  $\delta\psi\sim r^\alpha$ with 
  \begin{align}\label{char2}
  \alpha^2+3\alpha+4=0~~,~~\alpha=-\frac{3}{2}\pm\frac{i}{2}\sqrt{7},
 \end{align}
  and hence it is unstable there.  We must search for a stable solution.
  
  Since the term  on the right hand side of equation\ (\ref{eomnogauge})  diverges
  as $\psi\rightarrow0$ or $\psi\rightarrow \pi$ we also have the possibility of a gapped (Minkowski) embedding. 
  We thus expect this one to
  be the stable solution.
  This is confirmed by
  the numerical solution of the equation~(\ref{eomnogauge}), the details of which we will discuss in Section 4. However, in our numerical investigations, we find that
  the Minkowski embedding is only slightly favoured
  compared to the constant one as its energy splitting is significantly smaller than the value of the ultraviolet cutoff, $1/R$. 
  
  \subsection{Introducing charge density and magnetic field \label{eom}}
  
  Now, we will introduce a charge density and a magnetic field by introducing a gauge field
\begin{align}
2\pi\alpha'{ F}=\sqrt{\lambda}\alpha'\left[
 \frac{d}{dr}a(r)dr\wedge dt +b\,dx\wedge dy\right].
\end{align}
The parameter $b$ is a constant and it is related to the magnetic field, $B$.  The function $\frac{d}{dr}a(r)\equiv a'(r)$ is related 
to the world-volume electric field which is necessary in order to introduce a 
uniform charge density, $\rho$.  
They are related by the following identities,
\begin{equation}
b=\frac{2\pi}{\sqrt{\lambda}}B, \hspace{0.5cm} \rho=\frac{1}{V_{2+1} }\,\frac{2\pi}{\sqrt{\lambda}}\,\frac{\delta S}{\delta a'}.
\end{equation}
  The action for the D7 brane (\ref{sd7})  becomes
  \begin{align}
  S_{7}&=-{\cal N}_7 \int_0^{\infty} dr {\cal L}_7\nonumber\\
  &=-{\cal N}_7 \int_0^\infty dr \left[ f^{\frac{1}{2}}\sin^4\psi(r)\sqrt{(r^4+fb^2)(1+(r\psi'(r))^2-(a')^2)}
  \right.  \left.+4a'(r) b c(\psi)
  \right], \nonumber
  \end{align}
  where ${\cal N}_7$ was given in equation\ (\ref{Snorm}).
 The cyclic variable $a(r)$ can be eliminated by a Legendre transform in favour of an integration constant, $q$, proportional to the charge
 density
 \begin{equation}
 q=-\frac{\delta {\cal L}_7}{\delta a'}=\frac{3}{4}\, \frac{(2\pi)^3}{\sqrt{\lambda} N} \,\rho.
  \end{equation}
 This leads to the Routhian
 \begin{align}\label{r7}
  {\cal R}=-{\cal N}_7\int_0^\infty dr \,V(\psi,r)~\sqrt{1+(r\psi')^2},
 \end{align}
 where
  \begin{align}\label{pot}
  V(\psi,r)=   \sqrt{f\sin^8\psi(r^4+f\,b^2)+(q+4 b\,c(\psi))^2}. 
    \end{align}
 The equation of motion then reads
 $$
\frac{ \ddot\psi} {1+\dot\psi^2} +
\dot\psi \, \, \partial_y\ln V       = \partial_\psi \ln V,
  $$
 which becomes
  \begin{align}\label{general}
  \frac{\ddot\psi}{1+\dot\psi^2}&+
  \left[
  1+ 2r^4
  \frac{
  [1+2b^2R^4+2(1+b^2R^4)R^4r^4 ]\sin^8\psi
  }
  {
  f(r^4+fb^2)\sin^8\psi+(q+4b\,c)^2}
    \right]\dot\psi  \nonumber \\
  &    \qquad\qquad\qquad   
    -\frac{4\sin^7\psi\cos\psi f(r^4+fb^2) + 4b \,c'(q+4b\,c)}{ f(r^4+fb^2)\sin^8\psi+(q+4b\,c)^2}=0. 
  \end{align}
  We will discuss the solutions of this equation below.  We will begin with some simpler limiting cases.

  \subsubsection{Zero magnetic
  field  and non-zero density, $b\rightarrow 0, \,\, q\neq 0$}
  
 When we put $b\to0$, the equation of motion
  reduces to
  \begin{align}\label{eomnob}
  \frac{ \ddot\psi} {1+\dot\psi^2}  +\left[ 1+ \frac{   2r^4\sin^8\psi (1+2r^4R^4)}{ f\sin^8\psi \,r^4+q^2} \right]\dot\psi
  =\frac{4\sin^7\psi\cos\psi fr^4}{f\sin^8\psi\,  r^4+q^2}.
  \end{align}
  It is easy to argue that this equation will never have a gapped solution.  For such a solution to exist $\dot\psi$ has to
  diverge at the radius where the brane pinches off.   This divergence must be compensated by a divergence of the term on the
  right-hand-side of the equation.  But when $q$ is nonzero, this is impossible.  Therefore, we conclude that a charged brane must always
  reach the Poincare horizon of the black D3 brane at $r=0$.   
  
  Even if the D7 brane reaches the Poincare horizon, it is possible that parity is spontaneously broken.  In the dual field theory, this would
  correspond to a metallic phase of the system which still had a non-zero $\left< \bar\psi\psi\right>$ condensate.  We will find that this is not the case.
   We notice that the parity symmetric $\psi=\frac{\pi}{2}$ is a solution of  equation (\ref{eomnob}).  We can examine the stability of this
  constant solution by  linearizing about it.   Setting 
  $\psi=\frac{\pi}{2}-\delta\psi$, we get
  \begin{align}
  \delta\ddot{\psi}+ \left[1+\frac{2 r^4 (1+2r^4 R^4)}{r^4(1+R^4r^4)+q^2}
  \right] \delta\dot\psi+\frac{4r^4(1+r^4 R^4)}{r^4(1+r^4 R^4)+q^2}\,\delta\psi=0.
  \end{align}
  For large $r$ we get again the characteristic equation~(\ref{char1}) whereas for small $r$ the equation reduces to
  $\delta\ddot{\psi}+\delta\dot\psi=0$.  This equation has  a regular un-normalizable and normalizable solution,
  $\psi(r)\sim c_1+c_2/r$ and this solution is stable, at least under small fluctuations.    
  Hence, there is no sign of an instability which fits the picture that there is no
  solution for the gauge field with an $r$-dependent $\psi$ for $q\neq 0$ unless $b$ is also non-vanishing, cf.\ section~\ref{symmetries}.
  
  \subsubsection{Magnetic field, charge neutral case, $b\neq 0$, $\rho=0$}
  
 Next, let us consider the case $q= 0, \,\, b\neq 0$.  In this case the equation of motion reduces to
  \begin{align}
  \frac{\ddot\psi}{1+\dot\psi^2}+
  \left[
  1+ 2r^4
  \frac{
  [1+2b^2R^4+2(1+b^2R^4)R^4r^4 ]\sin^8\psi
  }
  {
  f(r^4+fb^2)\sin^8\psi+(4b\,c)^2}
    \right]\dot\psi\qquad \qquad\qquad&  \nonumber \\
 -\frac{4\sin^7\psi\cos\psi f(r^4+fb^2) +16 \,b^2c\,c'}{ f(r^4+fb^2)\sin^8\psi+(4b\,c)^2}=0. 
 \label{q=0} 
 \end{align}
 Again $\psi=\frac{\pi}{2}$ is a solution since $c(\pi/2)=0$.  To check for stability,  we
 linearize about the constant solution to get  
 \begin{align}
 \delta\ddot{\psi}+ \left[1+
 2r^4
  \frac{[1+2b^2R^4+2(1+b^2R^4)R^4r^4 ]
  }
  { f(r^4+fb^2)}
    \right]\delta\dot\psi
    +\frac{4f(r^4+fb^2) -16\,b^2 }{ f(r^4+fb^2)}\delta\psi=0. \nonumber
 \end{align}
 For large $r$ we recover once more the characteristic equation~(\ref{char1}). For small $r$ we get
  \begin{equation}
\delta \ddot{\psi}+\delta\dot\psi-12\delta \psi=0.
  \end{equation}
This is solved by 
 $\delta\psi\sim r^\alpha$ with 
  \begin{align}\label{char3}
  \alpha^2+\alpha-12=0~~,~~\alpha=-\frac{1}{2}\pm\frac{7}{2},
 \end{align}
 which are the required behaviours of normalizable and un-normalizable solutions. 
 Hence, as in the case of zero field and nonzero density, this case of nonzero field at zero density seems to have
 a stable constant solution.   Our numerical investigations did not reveal other solutions which is in accordance with
 the symmetry considerations in section~\ref{symmetries}. 
   Remember that, at vanishing density, and without a magnetic
 field, the solution is gapped.  As soon as we turn on a small magnetic field the gap  goes to zero.  
 This is an instance of ``{\it anti-catalysis}'' of symmetry breaking by a magnetic field, where turning on the
 field restores the symmetry.

  \subsubsection{Magnetic field and density, $b\neq 0~,~q\neq 0$}
  
  Let us now return to the most general equation~(\ref{general}).  We introduce the filling fraction $\nu$ by
 \begin{align}
 \nu= \frac{2\pi}{N}\frac{\rho}{B}= \frac{1}{\pi}\,\frac{2}{3}\,\frac{q}{b}.
 \end{align}
 Furthermore,  we rescale our variables in the following way
 \begin{align}
 \tilde{r}=r/\sqrt{b},\hspace{0.5cm} \tilde{R}=R\sqrt{b}.
\end{align}
Then the equation reads 
 \begin{align}\label{generalrescaled}
  \frac{\ddot\psi}{1+\dot\psi^2}+&
  \left[  1+ 2\tilde{r}^4
  \frac{
  [1+2\tilde{R}^4+2(1+\tilde{R}^4)\tilde{R}^4\tilde{r}^4 ] \sin^8\psi }
  { \tilde{f}(\tilde{r}^4+\tilde{f})\sin^8\psi+\frac{9}{4}(\pi\nu+\beta)^2}
    \right]\dot\psi  \nonumber \\
 &   \qquad\qquad\qquad   
    -\frac{4\sin^7\psi\cos\psi \tilde{f}(\tilde{r}^4+\tilde{f}) + \frac{9}{4}\,\beta'(\pi\nu+\beta)}{ \tilde{f}(\tilde{r}^4+\tilde{f})\sin^8\psi+
 \frac{9}{4}   (\pi\nu+\beta)^2}=0,
  \end{align}
where
$\tilde{f}=(1+\tilde{r}^4\tilde{R}^4)$ and where
\begin{equation}
\beta(\psi)=\frac{8}{3}\,c(\psi).
\end{equation}
We notice that this equation is invariant under the transformation
\begin{equation}
\nu\rightarrow -\nu, \hspace{0.5cm} \psi\rightarrow \pi-\psi.
\end{equation}
Hence we can restrict ourselves to considering only $\nu\geq 0$. Furthermore, we notice that under the
rescaling the Routhian turns into
\begin{align}
{\cal R}=-{\cal N}_7\,  b^{3/2} \int_0^{\infty} d\tilde{r}\sqrt{\tilde{f}\sin^8\psi (\tilde{r}^4+\tilde{f})+
{{9}\over{4}}\,\left(\pi\nu+\beta(\psi)\right)^2}\,\cdot 
\sqrt{1+(\dot\psi)^2}.
\end{align}
 For large $\tilde{r}$ equation~(\ref{generalrescaled}) simplifies to
 \begin{equation}\label{larger}
   \frac{\ddot\psi}{1+\dot\psi^2}+
5\dot\psi 
  -\cot\psi  =0,
 \end{equation}
 which is the same as the large $r$ version of the equation of motion when no gauge fields are present, cf.\ equation\ (\ref{eomnogauge}).
   At small $r$ we get 
$$
  \frac{\ddot\psi}{1+\dot\psi^2}+
  \dot\psi -\frac{4\sin^7\psi\cos\psi  +\frac{9}{4}\beta'(\pi \nu+\beta)}{ \sin^8\psi+\frac{9}{4}(\pi\nu+\beta)^2}=0.
  $$
  The possible asymptotic values for $\psi$ as $\tilde{r} \rightarrow 0 $ are 
 given by the zeros
  of the numerator of the last term above, i.e. 
 these are $\psi=0$, $\psi=\pi$ or the $\psi$'s which fulfill
\begin{align}
   &4\sin^3\psi\cos\psi  + \frac{9}{4}\cdot\frac{8}{3}\,(\pi\nu+\beta(\psi))=\nonumber \\
-&6\cos\psi \sin\psi +6\left(\psi+\pi\left(\nu-1/2\right) \right)=  \label{potzero} 0.
\end{align}
 In particular, we note, that $\psi=\frac{\pi}{2}$ is not a solution except in the special case $\nu\propto q=0$, treated above. 

We observe that, as expected, it is possible to have Minkowski embeddings for particular values of $\nu$. These are the
values of $\nu$ for which the denominators in equation\ (\ref{general}) and~(\ref{generalrescaled}) go to zero as $\psi\rightarrow 0$ or as $\psi\rightarrow \pi$
which are
\begin{equation}
\nu=\pm \frac{1}{2}.
\end{equation}

It is easy to show that for $\nu>\frac{1}{2}$ the equation~(\ref{potzero}) does not have any zeros whereas for
$0<\nu< \frac{1}{2}$ it does.  The zero moves from $\psi=\frac{\pi}{2}$ at $\nu=0$ to $\psi=0$ at $\nu=1/2$.
Furthermore, one can show, that the derivative of (\ref{potzero}) with respect to $\psi$ is always positive
at the zero. Accordingly, 
we expect that for $\nu>\frac{1}{2}$ the angle $\psi$ tends to $0$ for $r\rightarrow 0$ whereas for   
$\nu\in \left[0,\frac{1}{2}\right]$ the angle $\psi$ tends to the zero of equation~(\ref{potzero}). As mentioned earlier, the equation~(\ref{generalrescaled}) is invariant under $\nu\leftrightarrow -\nu$ and 
 $\psi\leftrightarrow \pi-\psi$.

\section{Numerical results\label{numerical}}

In our numerical investigations we set $R=1$ and consider the system for various values of $b$ and $\nu$ (or eqivalently $q$).
For all values of our parameters, the large $r$ version of the equation of motion, i.e.\ (\ref{larger}), permits the 
solution
\begin{equation}
\psi(r)=\frac{\pi}{2} -\frac{c_1}{r}-\frac{c_2}{r^4}-\ldots
\end{equation}
where the constant $c_1$ has the interpretation as the fermion mass in the dual gauge theory. We always consider
the massless case, $c_1=0$.
In cases where there are competing solutions (which turns out to be only for $\rho=b=0$) we compare their energies (i.e.\ minus the Routhians) to determine
which one is the preferred one.

\subsection{Vanishing magnetic field}

For $q=b=0$ we find, in accordance with our expectations (cf.\ section~\ref{nogaugefields}), that there exists
a Minkowski embedding and that this one is preferred in comparison with the trivial embedding. 
The preferred solution is the one with the lowest value
of $-S_{DBI}$ (which is equal to minus the Routhian for $q=b=0$). The energy of the gapped solution minus the
energy of the constant solution is
 $\Delta E=-2.8 \cdot 10^{-3}\, {\cal N}_7/R^3$. which is numerically much smaller than the ultra-violet cut-off.
We determine it by using numerical integration in the interval $\log (r) \in [-\infty,1.5]$ and approximating the contribution from
the interval $[1.5,\infty]$ by using a series expansion for the solution in powers of $\frac{1}{r}$ up to and including terms of order $1/r^{28}$. (We compare the energy of gapped and constant solutions for $b\neq 0$ in section~\ref{gappednu}, see in particular
figure~\ref{Gappedvsconstant}.)
\begin{figure}
\begin{center}
\includegraphics[scale=0.7]{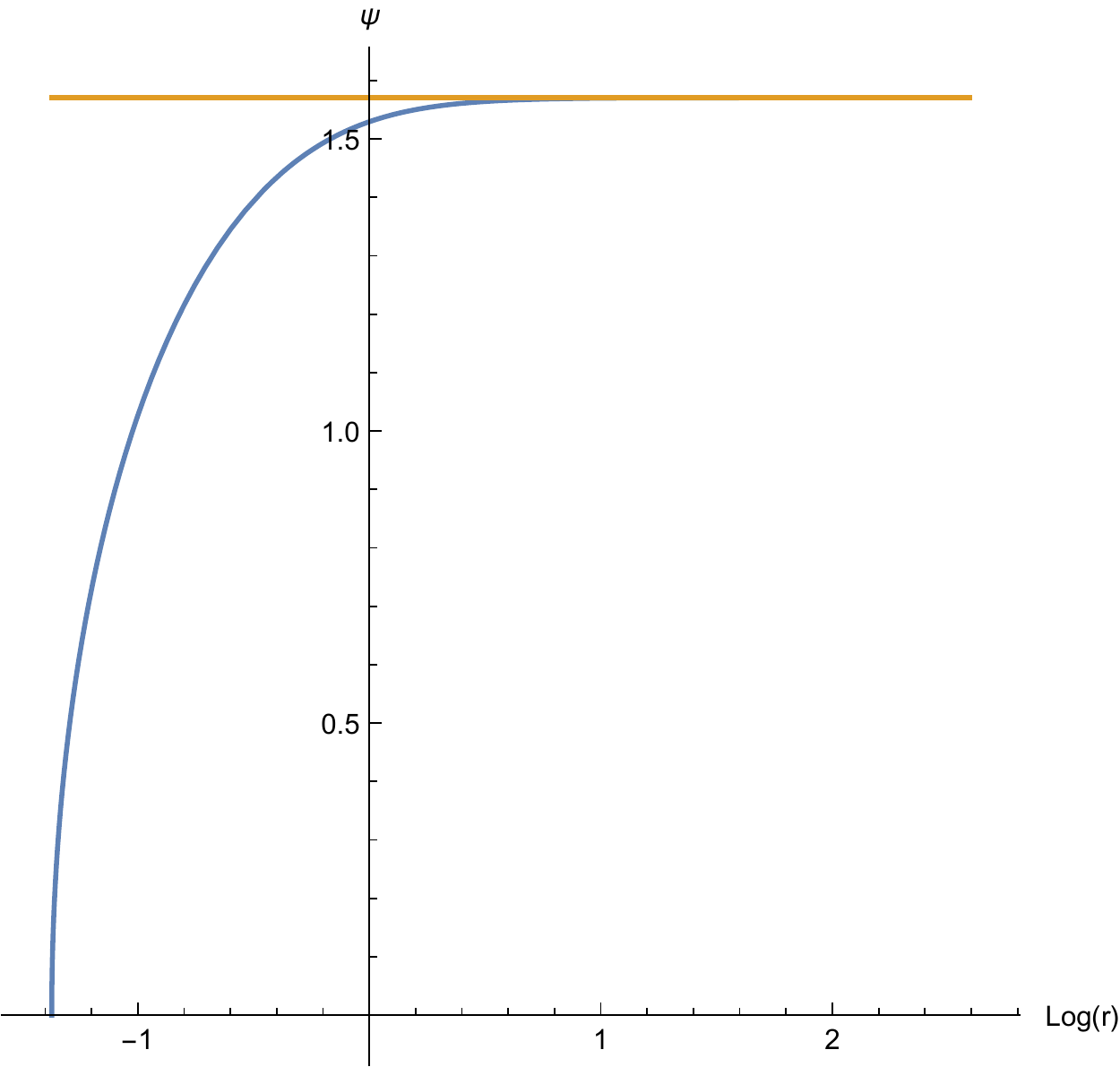}
\end{center}
\caption{The gapped and the constant solution for $q=b=0$.\label{QzeroBzero}}
\end{figure}
For $\nu\neq 0$, $b=0$, it follows from our analysis in section~\ref{symmetries} that only the constant solution is compatible with the symmetries of the system. This is confirmed by the numerical investigations. Notice that for $\nu=0$, $b\neq 0$ we also
only find the constant solution. This case is discussed in more detail in the subsequent sections.

\subsection{Ungapped solutions for $b\neq 0$}   
\begin{figure}
\begin{center}
\includegraphics[scale=.9]{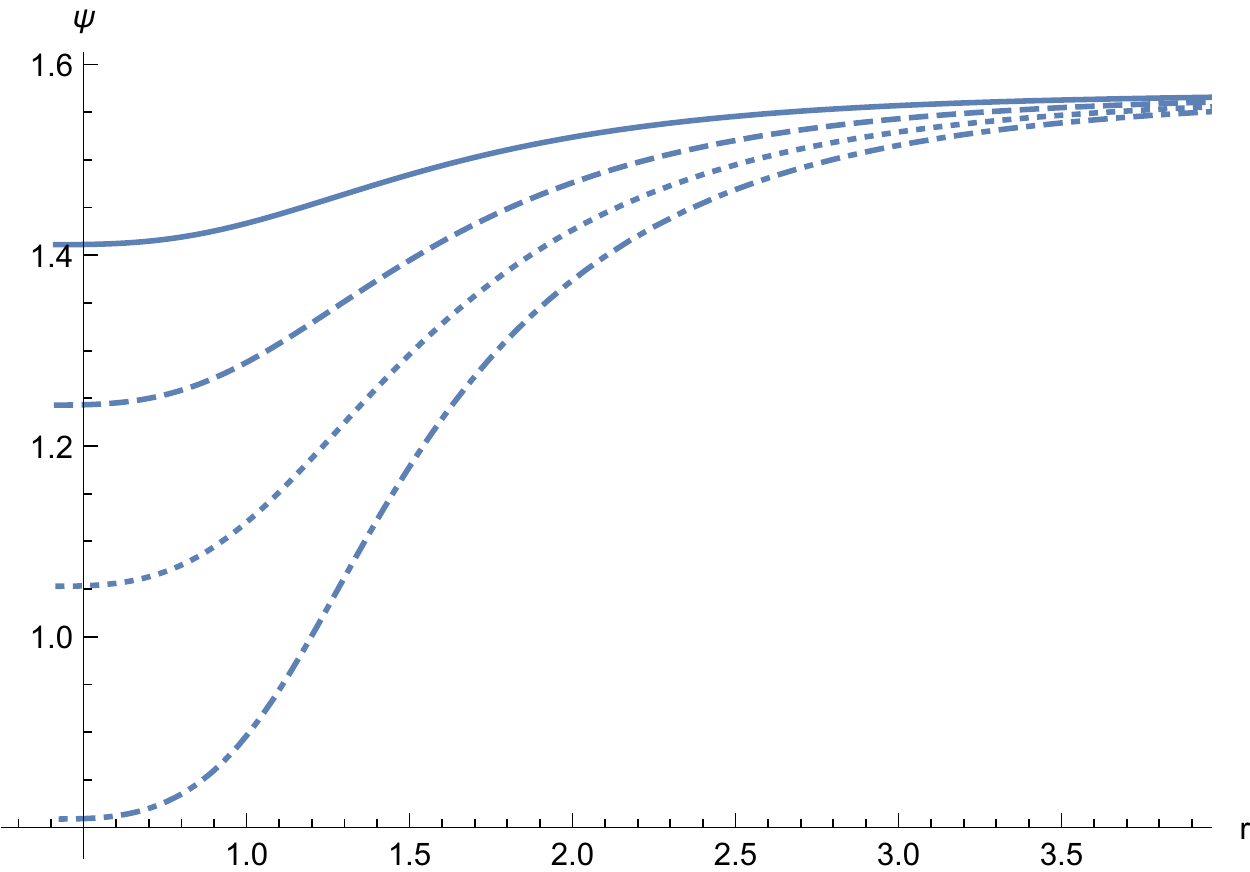}
\end{center}
\caption{The angle $\psi$ as a function of $r$ for $b=0.15$ and $\nu=0.1$ (full line), $\nu=0.2$ (dashed), $\nu=0.3$ (dotted)
and $\nu=0.4$ (dotdashed).\label{nulessthan05}}
\end{figure}

\begin{figure}
\begin{center}
\includegraphics[scale=.9]{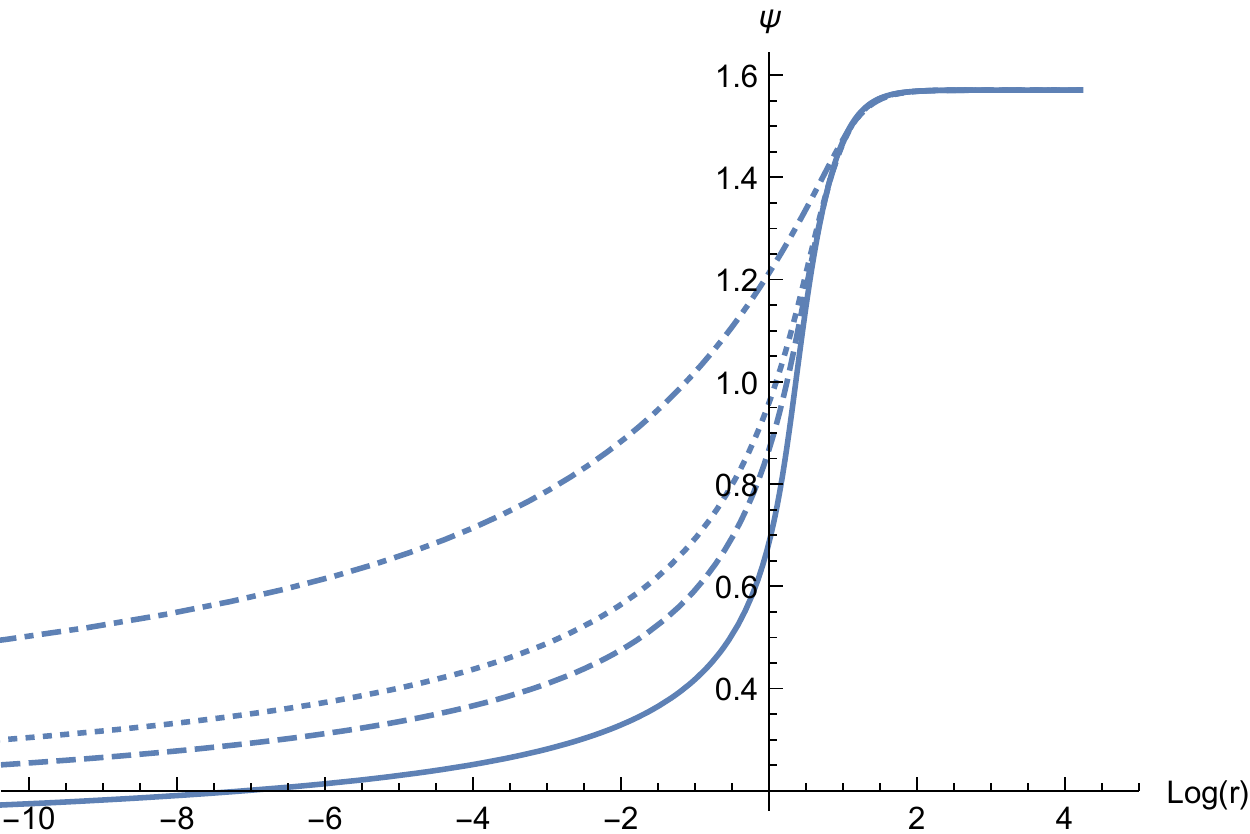}
\end{center}
\caption{The angle $\psi$ as a function of $\log(r)$ for $b=0.15$ and $\nu=0.6$ (full line), $\nu=0.8$ (dashed), $\nu=1.0$ (dotted)
and $\nu=2.5$ (dotdashed).\label{nulargerthan05vertical}}
\end{figure}

\begin{figure}
\begin{center}
\includegraphics[scale=.9]{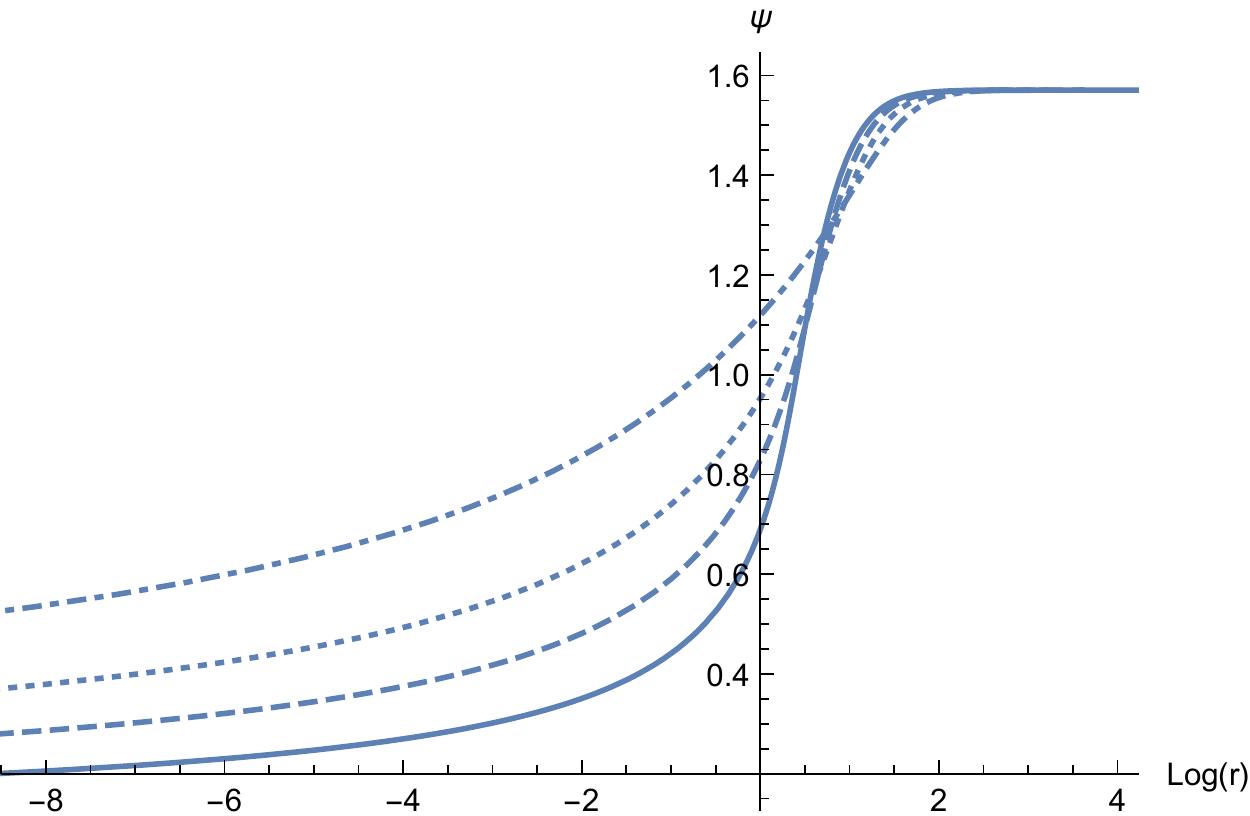}
\end{center}
\caption{The angle $\psi$ as a function of $\log(r)$ for $q=9/80\, \pi$ and $\nu=5/8 $ (full line), $\nu=5/6$ (dashed), $\nu=5/4$ (dotted)
and $\nu=5/2$ (dotdashed).\label{nulargerthan05horizontal}}
\end{figure}

As argued in section~\ref{eom}, gapped solutions are only possible for $\nu=\pm 1/2$ or for 
$q=b=0$. For all other choices of parameters the solutions must be un-gapped. Furthermore, the
constant solution is only a possible solution for $\nu=0$ (assuming $b\neq 0$).  We do numerically
find un-gapped solutions for all $\nu\in ]0,1/2[$ and $\nu> 1/2$ and no competing solutions seem to exist.
In accordance with our expectations, for
$\nu\in ]0,1/2[$ the angle $\psi$ tends to the zero of equation~(\ref{potzero}) for $r\rightarrow 0$. Notice that this zero
is $\psi=\frac{\pi}{2}$ for $\nu\rightarrow 0$.
The behaviour
of the angle $\psi$ as a function of $r$ is shown in figure~\ref{nulessthan05} for $\nu=0.1, 0.2, 0.3, 0.4$  and for
$b=0.15$. (This corresponds to following a vertical line below the $\nu=\frac{1}{2}$ curve in the $(b,q)$ diagram.)
For $\nu> 1/2$ the angle $\psi$ tends to zero as $r \rightarrow 0$, (albeit very slowly). 
In figure~\ref{nulargerthan05vertical} we show the angle $\psi$ as a function of $\log(r)$ for $\nu=0.6,0.8,1.0,2.5$ and
for $b=0.15$. (This corresponds to moving vertically above the line $\nu=\frac{1}{2}$ in the $(b,q)$-plane).
Furthermore, in figure~\ref{nulargerthan05horizontal} we show the angle $\psi$ as a function of $\log(r)$ for $\nu=5/8,5/6,5/4,5/2$ and
for $q=9/80\,\pi$. (This corresponds to moving horisontally in the $(b,q)$ plane in the region $\nu>\frac{1}{2}$  along a
line which intersects the $\nu=\frac{1}{2}$ line in the point $b=0.15$.).

\subsection{Gapped solutions for $b\neq 0$\label{gappednu}}

\begin{figure}
\begin{center}
\includegraphics[scale=.8]{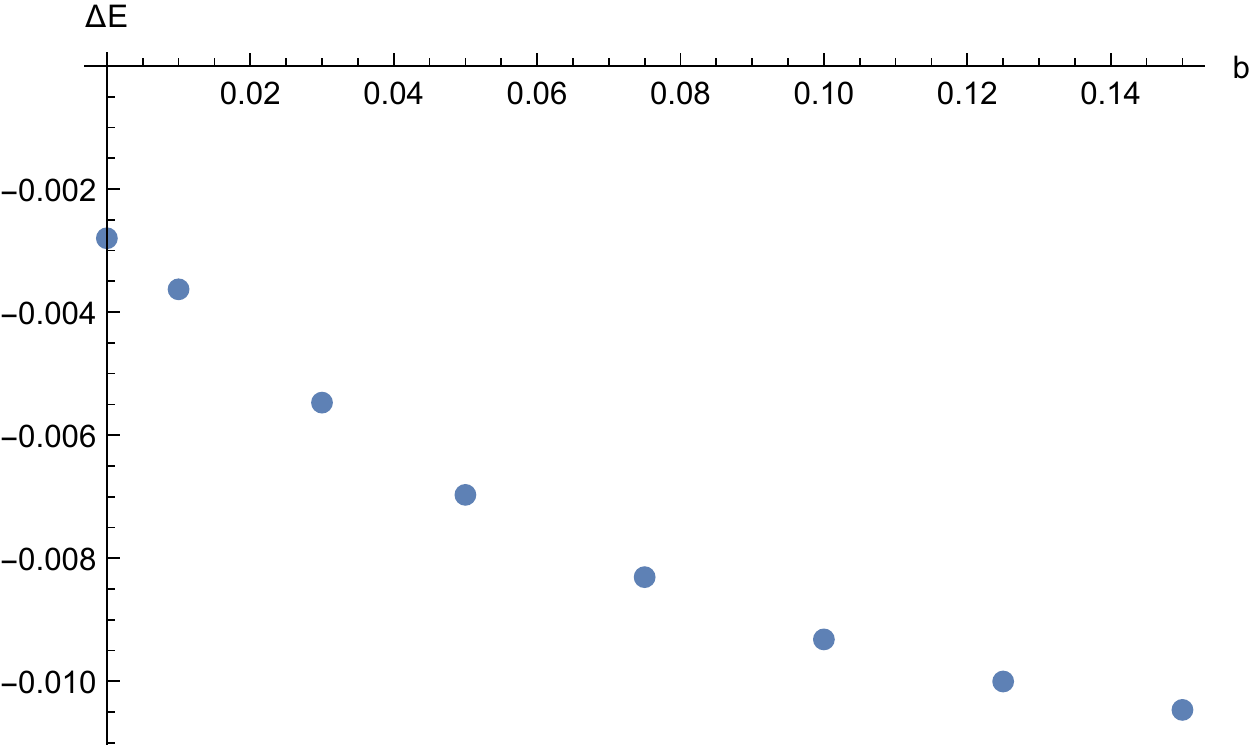}
\end{center}
\caption{The energy of the gapped solution for $\nu=\frac{1}{2}$ minus the energy of the constant
solution for $\nu=0$ as a function of $b$.\label{Gappedvsconstant}}
\end{figure}

As shown in section~\ref{eom} the system allows for solutions of Minkowski type, i.e. gapped solutions when
$\nu=\pm1/2$.  These gapped solutions are indeed found
numerically. No competing solutions are found. In particular, as noted above, the constant $\psi=\frac{\pi}{2}$ does
not solve the equation of motion. In figure~\ref{Gappedvsconstant} we show the energy of the gapped solution for 
$\nu=\frac{1}{2}$  minus the energy of the constant solution for $\nu=0$ as a functions of $b$. Notice that only differences
between energies are finite. The figure shows that for a given $b$ the system with $\nu=\frac{1}{2}$ has a lower engy
than the system with $\nu=0$. This continues to be the case as $b\rightarrow 0$ and is consistent with the result 
reported above that the gapped solution is favored over the constant one for  
$q=b=0$.

\begin{figure}
\begin{center}
\includegraphics[scale=0.9]{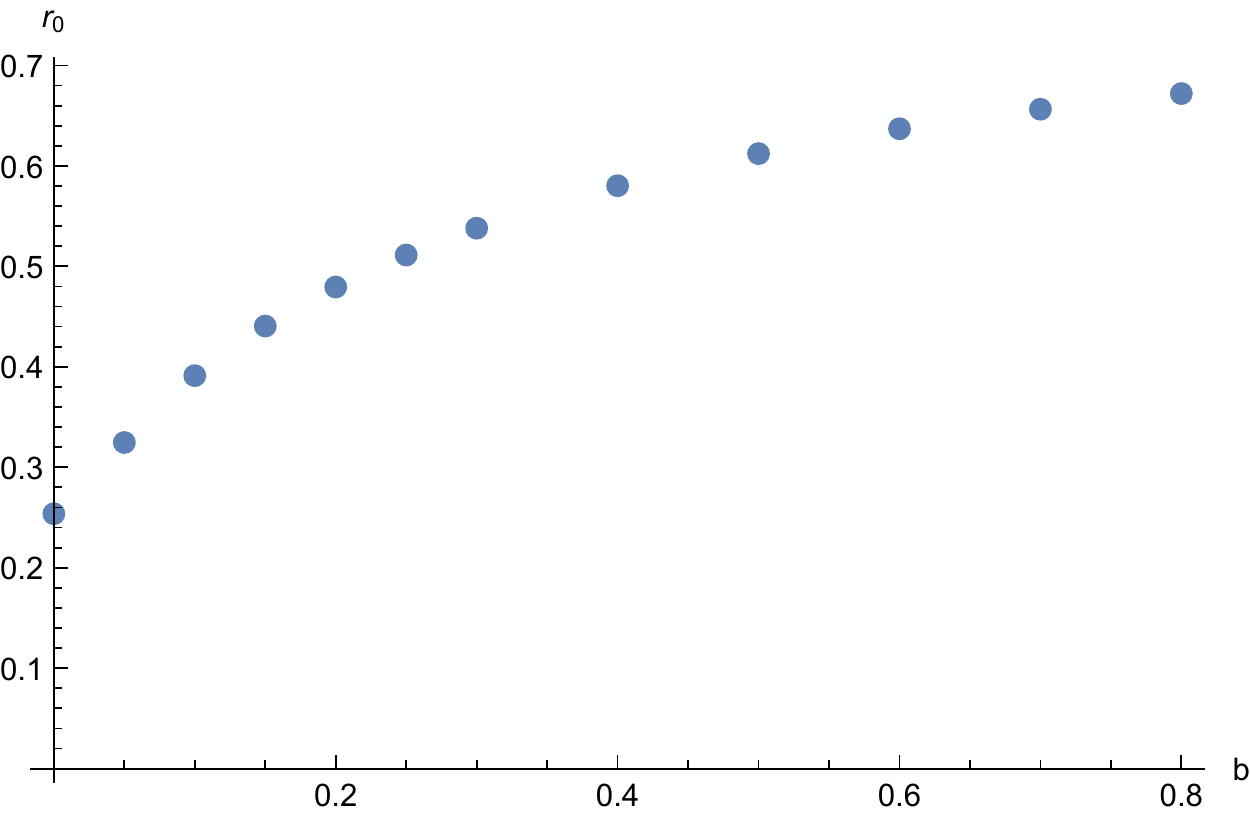}
\end{center}
\caption{The cap-off radius, $r_0$, as a function of the magnetic field, $b$. Fitting to a functional form
of the type $r_0=k_0+k_1 \,b^\alpha$ one gets $k_0\approx 0.24$,  $k_1\approx  0.50$, $\alpha\approx 0.49$.\label{r0vsb}}
\end{figure}

For the gapped solutions it is in addition of interest to study how the gap depends on $b$. 
For symmetry
reasons we consider only $\nu=1/2$ in the following. In figure~\ref{r0vsb}
we show the value of $r_0$ versus $b$ where $r_0$ is the radius where the D7 brane caps off. 
Fitting to a functional form
of the type $r_0=k_0+k_1 \,b^\alpha$ one gets $k_0\approx 0.24$,  $k_1\approx  0.50$, $\alpha\approx 0.49$.

\section{Conclusion\label{States}}

In this work, we have   confirmed anomaly matching between the large coupling and weak coupling limits where,
in both cases, an insulating state of the system has a zero-field Hall conductivity and the perturbative and strong coupling values of this conductivity are identical.   Secondly, we have examined the nature of the solutions of the cut-off, strongly coupled limit, 
particularly the issue of whether they are gapped or un-gapped, in the presence of a magnetic field and charge density.  
The phase diagram of the D3-probe-D7 theory is summarized in figure \ref{fig1}. The only point in that diagram where we find competing solutions
and have to decide on the phase is at the origin where the $\nu=\frac{1}{2}$ and the $\nu=-\frac{1}{2}$ solutions are degenerate, and are related by parity, P, 
or by CP. Everywhere else in the diagram, as far as we have investigated, the solution seems to be unique. 

Now, we can imagine the following scenario.   We begin with the D7 brane with a charge density $\rho$ and non-zero magnetic field, tuned so that  $\nu=\frac{1}{2}$.  It is an incompressible  charge-gapped state with $\frac{N}{2}$ units of Hall conductivity.  We then begin to decrease the filling fraction by decreasing the charge density $\rho$,
while holding the magnetic field $B$ fixed. In doing this, we immediately move away from the incompressible state.  The charge gap goes to zero and the system is metallic,
but it still has a Hall conductivity as well as, because it is now metallic,  a non-zero longitudinal conductivity.   The compressibility is due to the fact that when we move away from $\nu=\frac{1}{2}$, the D7 brane can no longer have a charge-gapped embedding.  The world volume of the D7 brane must have a spike which goes to the Poincare horizon of $AdS_5$.   We do still expect
(and confirm by numerical computations) that $\psi(r)$ is a non-constant function. This is interpreted as a non-zero chiral condensate. Of course, this is consistent
with the fact that we are in a region where $B$ and $\rho$ are non-zero, therefore $P$, $C$ and $CP$ are all explicitly broken and a chiral condensate is 
not suppressed by symmetry.  
As we continue to lower the density $\rho$ toward zero, the magnitude of the 
chiral condensate decreases and it eventually reaches zero when $\rho$ reaches zero. This behaviour is such that $\psi(r)$ seems to go from a smooth
weakly $r$-dependent function to a constant $\psi=\frac{\pi}{2}$ as $\rho$ approaches zero.  This is a result of the boundary condition that, at the lower extremity
of the D7 brane, $\psi(0)$ must satisfy equation (\ref  {potzero}) which has a solution which smoothly approaches $\frac{\pi}{2}$ as $\rho$ approaches zero.  
 This is a point with $CP$ symmetry ($B$ is non-zero,
$\rho$ is zero).
  It is interesting to ask whether the behaviour at this point is a quantum phase transition or if it is a completely smooth crossover between the two phases.
  Our numerical investigations so far are not accurate enough to resolve this issue.  An interesting quantity to compute would be the exponent  $\alpha$ in
  $$
  \left<\bar\psi\psi\right> \sim q^\alpha,~~~~{\rm as}~q\to 0.$$
  Then, in the new state, as we move $\nu$ below zero, toward \mbox{$\nu=-\frac{1}{2}$}, the spike that connects the D7 worldvolume to the Poincare horizon 
shrinks again and it disappears at $\nu=-\frac{1}{2}$ where the D7 brane   is charge-gapped again.  The result is that, for the same value of $B$, 
the D7 brane has two incompressible Hall states, with $\nu=\pm\frac{1}{2}$ and a possible phase transition at  $\nu=0$.    Although it is not strictly a transition between Hall plateaus, which are absent in this model, this is the closest that we can come to 
the plateau transition which was the objective of the investigations in references \cite{Davis:2008nv}   
 and  \cite{Mezzalira:2015vzn} and it deserves further investigation.
 
 We could also imagine doing what we describe in the above paragraph, but holding $\rho$ fixed and varying $B$.  Now, we are in the $|\nu|>\frac{1}{2}$ regime, so
 for ungapped states, the angle $\psi(r)$ must go to zero at the horizon for all states with the exception of the one where $B=0$ where it should be a constant $\psi=\frac{\pi}{2}$.  For this reason, we expect that
 the behaviour of solutions as $B\to 0$ as being (perhaps mildly) non-analytic.  Again, as $B\to 0$, the chiral
 condensate should also approach zero and $$
  \left<\bar\psi\psi\right> \sim B^{\tilde\alpha},~~~~{\rm as}~B\to 0.$$
  And, again, our numerical computations are not precise enough to resolve the exponent $\tilde\alpha$.

Some recent experiments on integer Hall states in graphene have measured the magnetic field dependence of
the energy gap which separates Hall plateaux \cite{kimetal,geimetal}.  For the most part these scale as $\Delta\sim B$.  We have computed
the $B$-dependence of the gap in the $\nu=\frac{1}{2}$ state.  Our data, which is plotted in figure \ref{r0vsb} gives that
$\Delta \sim |B|^{\frac{1}{2}}$ which differs significantly from experimental results.

\acknowledgments

The work of G.W.S.~is supported in part by NSERC of Canada. G.W.S.~acknowledges the kind hospitality of 
the Niels Bohr Institute, NORDITA and BIRS during the course of this work.
C.K.\ was supported by FNU through grant number DFF -- 1323 -- 00082. C.K.\ would like to thank the
University of British Columbia, BIRS and NORDITA for their kind hospitality.


\end{document}